\documentclass[
showpacs,
floatfix,
aps,
pra,
amsmath,
twocolumn,
groupaddress,
]{revtex4-1}
\usepackage{graphicx}  
\usepackage{dcolumn}   
\usepackage{bm}        
\usepackage{amssymb}   
\usepackage{amsfonts}
\usepackage{verbatim}
\usepackage{epstopdf}
\DeclareGraphicsExtensions{.eps}

\usepackage{amsmath}
\usepackage{amsfonts}
\usepackage{ulem}
\usepackage{color}
\usepackage{mathtools}
\usepackage{natbib}

\newcommand{\eins}{\mbox{$1 \hspace{-1.0mm} {\bf l}$}}

\newcommand{\ket}[1]{|#1\rangle}

\def\lsim{\mathrel{\rlap{\lower4pt\hbox{\hskip1pt$\sim$}}
    \raise1pt\hbox{$<$}}}                
\def\gsim{\mathrel{\rlap{\lower4pt\hbox{\hskip1pt$\sim$}}
    \raise1pt\hbox{$>$}}}                

\begin{document}
\normalem

\title{Accurate Qubit Control with Single Flux Quantum Pulses}

\author{R. McDermott}\email[Electronic address: ]{rfmcdermott@wisc.edu}
\author{M.G. Vavilov}
\affiliation{Department of Physics, University of
Wisconsin, Madison, Wisconsin 53706, USA}

\date{\today}

\begin{abstract}
We describe the coherent manipulation of harmonic oscillator and qubit modes using resonant trains of single flux quantum pulses in place of microwaves. We show that coherent rotations are obtained for pulse-to-pulse spacing equal to the period of the oscillator. We consider a protocol for preparing bright and dark harmonic oscillator pointer states. Next we analyze rotations of a two-state qubit system. We calculate gate errors due to timing jitter of the single flux quantum pulses and due to weak anharmonicity of the qubit. We show that gate fidelities in excess of 99.9\% are achievable for sequence lengths of order 20 ns.
\end{abstract}

\maketitle

\section{Introduction}

Josephson qubits are a leading candidate for scalable quantum information processing in the solid state \cite{Clarke08, Devoret13}. Gate and measurement fidelities are within reach of the threshold for fault-tolerant quantum computing based on topological surface codes \cite{Fowler12,Barends14}, and there is interest in scaling to larger multi-qubit circuits. A superconducting quantum computer that will outperform the best available classical machines will require thousands if not millions of physical qubits, and the wireup and control of a large-scale quantum processor presents a formidable technical challenge. It is highly desirable to integrate as much of the control and measurement circuitry as possible in the multi-qubit cryostat in order to reduce wiring heat load, latency, power consumption, and the overall system footprint. An obvious candidate for the cold control system is Single Flux Quantum (SFQ) digital logic, in which classical bits of information are stored in propagating fluxons, voltage pulses whose time integral equals the superconducting flux quantum $\Phi_0 = h/2e$ \cite{Likharev91, Bunyk01a}. There have been experimental demonstrations of SFQ-based circuits for qubit biasing \cite{Ohki05, Hassel06, Castellano07}, and fluxon-based schemes for qubit measurement have been proposed \cite{Fedorov07} and recently realized \cite{Fedorov14}. In addition, there has been a proposal to generate microwave pulses for qubit control by appropriately filtering SFQ pulse trains \cite{Takeuchi10}, although the required filter and matching sections would be challenging to realize practically. Up to now, however, there has been no compelling proposal for the realization of coherent quantum control of superconducting qubit and linear cavity modes by direct excitation \textit{via} SFQ pulses.

In this Article, we propose a scheme for the coherent control of qubit and linear cavity modes using resonant SFQ pulse trains. We demonstrate that SFQ-based gates are robust against leakage errors and timing jitter of the pulses, with achievable fidelities in excess of 99.9\% in gate times around 20 ns. In separate work we have analyzed a circuit quantum electrodynamics (cQED) measurement scheme wherein the qubit state is mapped to the binary digital output of a cryogenic microwave photon counter \cite{Govia14}. Taken together, these proposals point the direction for integration of a multi-qubit quantum processor with cold SFQ-based classical digital circuitry for both control and measurement.

This paper is organized as follows. In Section II, we describe the coherent control of harmonic oscillator modes and qubits with SFQ pulses, starting from a classical model and moving to a quantum description of the coupling Hamiltonian. This section includes our main numerical results for SFQ gate fidelity in the presence of pulse timing imperfections and higher energy levels of the qubit. Section III includes a detailed treatment of gate error due to finite SFQ pulse width, SFQ pulse timing jitter, and weak qubit anharmonicity. In Section IV we present our conclusions.

\section{Coherent Control via SFQ Pulses}

Control in superconducting qubits is typically accomplished \textit{via} shaped microwave pulses that realize arbitrary rotations over the Bloch sphere. Amplitude modulation of a resonant carrier wave concentrates drive power at the frequency of interest, and pulses are shaped to minimize power at nearby transition frequencies to avoid excitation out of the qubit manifold \cite{Motzoi09, Motzoi11}. We can gain intuition for the effectiveness of an arbitrary drive pulse at addressing a desired transition (or avoiding an undesired one) by considering a simple classical model of an $LC$ resonator. The drive waveform is coupled to the resonator from a time-dependent voltage source $V(t)$ through a coupling capacitance $C_c$ (see Fig. 1a). We find that the energy deposited in the resonator is given by
\begin{align}
E = \frac{\omega_0^2C_c^2}{2C'} \left|\widetilde{V}(\omega_0)\right|^2,
\label{eq:E}
\end{align}
where $C' = C + C_c$, $\omega_0 = 1/\sqrt{LC'}$, and where the tilde represents the Fourier transform $\widetilde{V}(\omega) = \int_{-\infty}^{\infty} V(t) \, e^{-i\omega t}\, dt$. The energy coupled to the resonator is proportional to the energy spectral density of the drive waveform at the resonator frequency.

Here we are interested in the response of a microwave resonator to an SFQ pulse. For state-of-the-art Nb-based SFQ technology, characteristic pulse amplitudes are 2~mV and pulse widths are around 1~ps. As the pulse widths are much less than the period of the microwave resonator, we can model the SFQ pulse as a Dirac $\delta$-function $V(t) = \Phi_0 \delta(t)$. In this case, we find $\widetilde{V}(\omega) = \Phi_0$ and Eq.~\eqref{eq:E} reduces to
\begin{align}
E_1 = \frac{\omega_0^2 C_c^2 \Phi_0^2}{2C'},
\label{eq:pulse}
\end{align}
where the subscript 1 indicates that we are referring to the response to a single pulse. Because the SFQ pulse width is much smaller than the oscillator period, the energy deposited is quite insensitive to the detailed shape of the SFQ pulse, and is determined rather by the time integral of the pulse, which is precisely quantized to a single flux quantum. For example, for a Gaussian SFQ pulse with standard deviation $\tau$, the above result is modified by the prefactor $e^{-\omega_0^2 \tau^2}$, which yields a correction of 0.02\% for $\tau$ = 0.5 ps and $\omega_0/2\pi$ = 5 GHz.

A single SFQ pulse produces a broadband excitation. For this reason, the single pulse is not useful for coherent manipulation of quantum circuits, since it doesn't offer the possibility to selectively excite individual transitions. The picture changes, however, when we consider driving the resonator with a train of SFQ pulses. The goal is to coherently excite the resonator by using a pulse-to-pulse separation that is matched to the resonator period. The approach is analogous to pumping up a swing by giving a short push once per cycle, as opposed to sinusoidally forcing the swing throughout the entire period of oscillation. We consider the driving voltage
\begin{align}
V_n(t) = \Phi_0\left[\delta(t) + \delta(t-T) + ... + \delta(t-(n-1)T)\right],
\end{align}
where $T$ is the separation between pulses and $n$ is the number of pulses. We find that the pulse train couples an energy to the resonator equal to
\begin{align}
\label{eq:En}
E_n = \frac{\omega_0^2 C_c^2 \Phi_0^2}{2C'}
\frac{\sin^2(n \omega_0 T/2)}{\sin^2(\omega_0 T/2)}.
\end{align}

It is worthwhile to consider the energy transferred by an SFQ pulse train to a typical cavity mode in a superconducting cQED circuit. We take $\omega_0/2\pi$~=~5~GHz, $C$~=~1~pF, and $C_c$~=~1~fF. We find that a single SFQ pulse couples only $6~\times~10^{-4}$ quanta to the cavity mode. However, for a resonant pulse train with $T$ equal to an integer multiple of cavity periods, the pulses add coherently, so that the total energy deposited in the cavity goes as $n^2$. Because of this quadratic scaling, only 40 pulses are required to populate the cavity with a single excitation, and this can be accomplished in the time $40 \times 2\pi/\omega_0$~=~8~ns. 

A recent proposal for cQED measurement based on microwave counting relies on the preparation of ``bright" and ``dark" cavity pointer states using a coherent drive pulse with length matched to the inverse detuning of the dressed cavity frequencies \cite{Govia14}. This protocol is readily adapted to SFQ excitation of the readout cavity. For a qubit-cavity system with dressed cavity resonances at $\omega_0 - \chi$ (or $\omega_0 + \chi$) corresponding to the qubit $\ket{0}$ (or $\ket{1}$) states, an SFQ pulse train with interval $T = 2\pi/(\omega_0 + \chi)$ and total number of pulses $n = (\omega_0 + \chi)/2\chi$ will coherently populate the cavity if the qubit is in the $\ket{1}$ state, while returning the cavity to the vacuum upon completion of the sequence if the qubit is in the $\ket{0}$ state.

\begin{figure}[t]
\includegraphics[width=.48\textwidth]{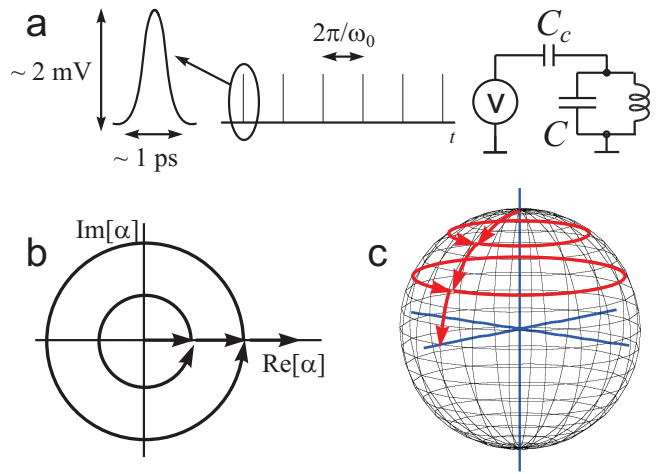}
\vspace*{-0.0in} \caption{(Color online) (a) Excitation of a resonator mode \textit{via} a train of SFQ pulses. The pulses are coupled to the resonator through the capacitance $C_c$. For Nb-based SFQ technology, pulse amplitudes are of order 2 mV and pulse widths of order 1 ps. (b) Trajectory in quadrature space for a cavity driven by a resonant SFQ pulse train. (c) Trajectory on the Bloch sphere for a qubit driven with a resonant SFQ pulse train.}
\label{fig:fig1}\end{figure}

Next we consider the response of the quantum oscillator to SFQ excitation. The time-dependent circuit Hamiltonian is written as:
\begin{align}
H = \frac{[\widehat{Q}-C_cV(t)]^2}{2C'} + \frac{\widehat{\Phi}^2}{2L}.
\end{align}
We decompose the Hamiltonian into the unperturbed free Hamiltonian $H_{\rm{free}}$ and a time-dependent excitation Hamiltonian $H_{\rm{SFQ}}$:
\begin{align}
H_{\rm{free}} &= \frac{\widehat{Q}^2}{2C'} + \frac{\widehat{\Phi}^2}{2L}
\nonumber\\
H_{\rm{SFQ}} &= - \frac{C_c}{C'} \, V(t) \widehat{Q}.
\end{align}
In terms of the usual raising and lowering operators, we have
\begin{align}
H_{\rm{free}} &= \hbar \omega_0 \hat{a}^\dag \hat{a},
\nonumber\\
H_{\rm{SFQ}} &= i C_cV(t) \sqrt{\frac{\hbar \omega_0}{2C'}} \left(\hat{a} - \hat{a}^\dag\right).
\end{align}
The effect of the SFQ pulse is to induce a coherent displacement of the cavity state by amount
\begin{align}
\alpha_{\rm{SFQ}} = -C_c\Phi_0\sqrt{\frac{\omega_0}{2\hbar C'}};
\end{align}
see Fig. \ref{fig:fig1}b.
The energy deposited by the pulse matches the classical expression (\ref{eq:pulse}). A sequence of $n$ pulses produces a coherent state with amplitude $\alpha_n=\alpha_{\rm SFQ}\sum_{k=0}^{n-1}\exp(-i k \omega_0 T)$ and mean energy $E_n=\hbar\omega_0|\alpha_n|^2$, consistent with the classical expression \eqref{eq:En}.

Next, we consider application of SFQ pulses to a two-level qubit. The Hamiltonian of the system becomes
\begin{align}
H_{\rm free} &= \frac{\hbar \omega_{10}}{2} \left(\eins - \hat{\sigma}_z \right),
\nonumber\\
H_{\rm SFQ} &= C_cV(t) \sqrt{\frac{\hbar \omega_{10}}{2C}} \hat{\sigma}_y,
\end{align}
where $\eins$ is the identity matrix and $\mathbf{\hat{\sigma}}$ are the usual Pauli matrices. We will work in the limit of a short, intense SFQ pulse that induces a discrete rotation of the state vector
about the $y$-axis by angle
\begin{align}
\delta \theta = C_c \Phi_0 \sqrt{\frac{2 \omega_{10}}{\hbar C}};
\label{eq:angle}
\end{align}
in between pulses, the qubit evolves under the influence of $H_{\rm free}$. (In Section III we consider the effect of finite SFQ pulse width and show that, for typical cQED frequencies, free evolution during the pulse can be safely neglected). The SFQ pulse train will induce coherent rotations when the free evolution periods are matched to the oscillation period $2 \pi/\omega_{10}$ of the qubit; see Fig. \ref{fig:fig1}c. For a qubit initially in state $\ket{0}$, the resonant pulse train yields a coherent rotation in the $xz$-plane. For a pulse interval that is slightly mismatched from the oscillation period, the state vector slowly drifts away from the $xz$-plane, and in the limit of a large timing mismatch the state vector undergoes small excursions about the north pole of the Bloch sphere.

As can be seen from Eq.~\eqref{eq:angle}, the angle of rotation induced by the SFQ pulse depends on the strength of the capacitive coupling to the qubit, which we take to be fixed. While tunable inductive couplers have been demonstrated \cite{Bialczak11}, it is unclear that they could be engineered to perform well on the picosecond timescales characteristic of the SFQ pulse. For that reason, it might prove necessary to work with a fixed rotation angle once the coupling to the qubit is determined by the circuit design. For small rotation angle $\delta\theta \sim 0.01$, the resulting gate error is at most $\delta \theta^2/4$. Moreover, this error can be further reduced by appropriately tailoring the timing delay between the SFQ pulses, but discussion of such sequences is beyond the scope of the current work.

Other potential sources of error in SFQ-based gates are timing jitter of the pulses and weak anharmonicity of the qubit. In Section III we provide a detailed analysis of these errors; here we summarize the main results. In the following we take as input states the six eigenstates of the Pauli operators, and we compute gate error as the state error averaged over these input states; this approach is equivalent to interleaved random benchmarking with single-qubit Clifford gates \cite{Barends14,Knill08,Chow09}.

The effect of a timing error $\delta t$ in the SFQ pulse is to induce a spurious rotation of the state vector by angle $\omega_{10} \, \delta t \sin \theta$, where $\theta$ is the instantaneous polar angle of the state vector. We assume that the arrival times of the individual pulses are distributed normally with standard deviation $\sigma$. To consider the effect of timing jitter on rotations derived from SFQ pulse trains, we need to specify the manner in which the SFQ circuit is clocked. If the pulse train is derived from a stable external frequency source (used, for example, to trigger a DC/SFQ converter \cite{Likharev91}), the timing jitter per pulse is independent of the length of the pulse train. Timing errors associated with each pulse are largely compensated by the following pulse, and error in the final pulse dominates error in the sequence as a whole.  Pulse timing jitter leads to the average gate error
\begin{figure}[t!]
\includegraphics[width=.48\textwidth]{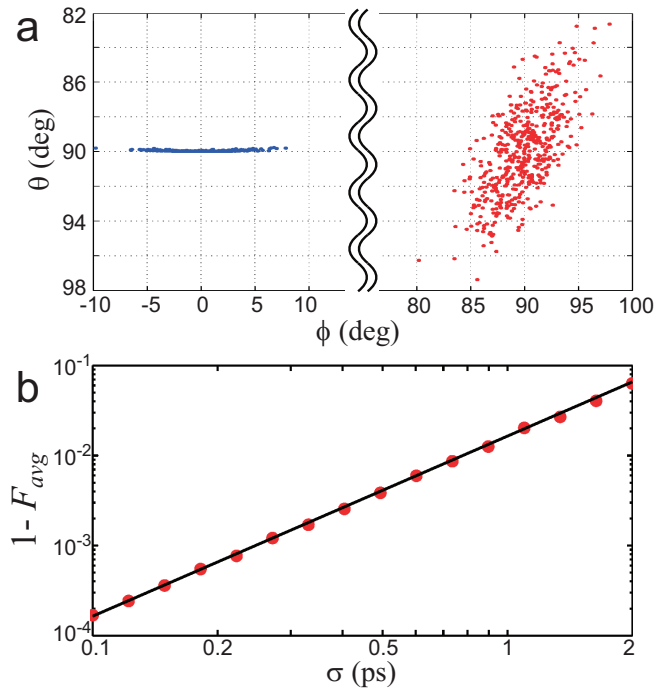}
\vspace*{-0.0in} \caption{(Color online) SFQ gate error due to timing jitter. (a) Scatter plot of polar $\theta$ and azimuthal $\phi$ angles of the Bloch vector following an SFQ-based $(\pi/2)_y$ rotation implemented in 100 pulses, for timing jitter $\sigma$ = 0.2 ps. The initial states are $\ket{0}$ (left) and $\left(\ket{0} + i \ket{1}\right)/\sqrt{2}$ (right), corresponding to target states $\left(\ket{0} + \ket{1}\right)/\sqrt{2}$ and $\left(\ket{0} + i \ket{1}\right)/\sqrt{2}$, respectively. (b) SFQ gate error \textit{versus} timing jitter $\sigma$ for an SFQ $(\pi/2)_y$ rotation implemented in 100 pulses. The points are the result of numerical simulations while the solid line is calculated from Eq. (\ref{eq:error}).}
\label{fig:fig2}\end{figure}
\begin{align}
1-F^{\rm ext}_{\rm avg} = \frac{(\omega_{10}\sigma)^2}{6}\left[\frac{\Theta^2}{n} + 1\right],
\end{align}
where the superscript ``ext" refers to the mode of clocking the SFQ pulse train from a stable external source. For practical purposes this timing jitter will introduce negligible gate error.

Next we consider the more demanding case where pulse timing errors accumulate incoherently, so that the timing jitter in the $n$th pulse is $\sqrt{n}$ larger than the timing jitter in the initial pulse. This could be the situation, for example, when the SFQ pulse train is generated internally from an SFQ clock ring. In this case, the deviation of the state vector from the desired trajectory grows as $\sqrt{n}$, leading to a degradation of gate fidelity that scales linearly with $n$. The timing jitter results in an average gate error

\begin{align}
1-F^{\rm int}_{\rm avg} = \frac{n(\omega_{10}\sigma)^2}{6},
\label{eq:error}
\end{align}
where the superscript ``int" refers to the internal clock used to generate the pulse train.

In the thermal regime, the timing jitter of the SFQ pulse scales as the square root of temperature \cite{Rylyakov99}, and average timing jitter per junction of 0.2 ps has been measured in a large-scale SFQ circuit operated at 4.2 K \cite{Bunyk01b}. For an SFQ circuit operated at reduced temperature in a dilution refrigerator, the timing jitter is expected to be lower, although quantum fluctuations will lead to nonnegligible jitter even for circuits operated at millikelvin temperatures. Moreover, if the SFQ pulse source is coupled to the qubit sample \textit{via} a long Josephson transmission line consisting of $N$ junctions, the qubit will see a $\sqrt{N}$ degradation of the timing jitter due to the sequential switching of the junctions in the line.

We have performed Monte Carlo simulations of gate error due to timing jitter for an SFQ $(\pi/2)_y$ rotation realized from 100 pulses, in the case where timing errors of the pulse generator accumulate incoherently, cf. Eq.~\eqref{eq:error}. The results are shown in Fig. \ref{fig:fig2}a-b. For the $\ket{0}$ state input, timing errors lead predominantly to $y$-errors. Small $z$-errors accumulate coherently and lead to a systematic underrotation of the state vector. For the input $(\ket{0} + i \ket{1})/\sqrt{2}$, which ideally is unaffected by the $(\pi/2)_y$ rotation, timing errors initially provide kicks in the $x$-direction; once $x$-errors are allowed to accumulate, subsequent SFQ pulses generate additional $z$-errors. In Fig. \ref{fig:fig2}b we show average gate error \textit{versus} pulse timing jitter $\sigma$. For $\sigma$ = 0.2 ps, we find average gate error of $6.6 \times 10^{-4}$.

A practical superconducting qubit is not an ideal two-level system \cite{Steffen03}. For a typical transmon qubit \cite{Koch07, Schreier08, Barends13}, the anharmonicity $(\omega_{10}-\omega_{21})/\omega_{10}$ is of order 4-5\%. A single strong SFQ pulse will induce a large spurious population of the $\ket{2}$ state as a result of its broad bandwidth, and leakage errors induced by fast SFQ control pulses have been considered previously \cite{Ohki07}. However, a resonant SFQ pulse train tailored to perform a desired rotation in the 0--1 subspace in a larger number of steps $n$ will show greatly reduced spectral density at $\omega_{21}$, enabling high-fidelity SFQ-based gates with acceptable leakage. We consider a three-level system with unpeturbed Hamiltonian
\begin{figure}[t]
\includegraphics[width=.495\textwidth]{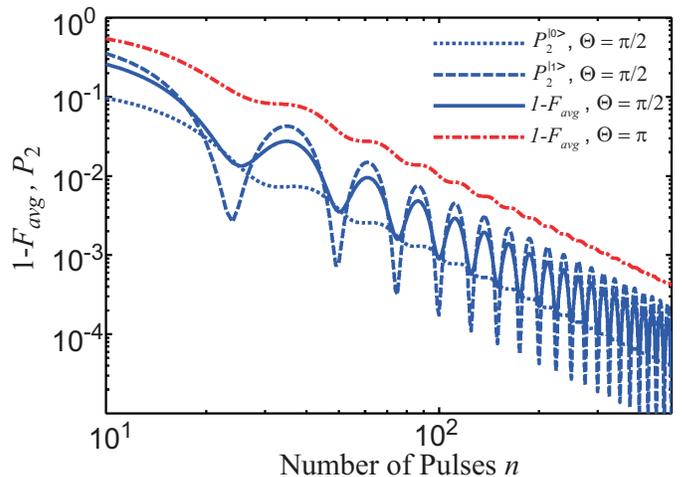}
\vspace*{-0.0in} \caption{(Color online) Average gate error and $\ket{2}$ state error $P_2$ for SFQ pulse trains \textit{versus} number of pulses $n$. Blue (lower) curves are for the SFQ implementation of the $(\pi/2)_y$ gate, and red (upper) curve is for the $\pi_y$ gate. Here, $\omega_{10}/2\pi$ = 5 GHz, $\omega_{21}/2\pi$ = 4.8 GHz, and gate error is computed as described in the main text.}
\label{fig:fig3}
\end{figure}
\begin{align}
H_{\rm free} = \begin{pmatrix*}[c]
0 &0 & 0 \\
0 & \hbar \omega_{10} & 0 \\
0 & 0 & \hbar(\omega_{10}+\omega_{21}) \end{pmatrix*}.
\end{align}
The charge induced on the qubit capacitance by the SFQ pulse leads to the Hamiltonian
\begin{align}
H_{\rm SFQ} = iC_c V(t) \sqrt{\frac{\hbar \omega_{10}}{2C}}
\begin{pmatrix*}[r]
0 &-1 & 0 \\
1 & 0 & -\sqrt{2} \\
0 & \sqrt{2} & 0 \end{pmatrix*}.
\end{align}
Here we consider the typical transmon parameters $\omega_{10}/2 \pi$~=~5~GHz and $\omega_{21}/2\pi$~=~4.8 GHz. We have examined gate fidelity and $\ket{2}$ state errors for resonant SFQ pulse trains designed to produce $(\pi/2)_y$ and $\pi_y$ rotations for a range of total numbers of pulses (and hence gate durations). In addition, we have computed the $\ket{2}$ state leakage $P_2^{\ket{j}}$ for the $(\pi/2)_y$ gate for initial qubit states $\ket{j} = \ket{0}, \ket{1}$. The results are shown in Fig. \ref{fig:fig3}. Gate error is dominated by leakage to the $\ket{2}$ state. Gate errors decrease as $n^{-2}$; by increasing the number of pulses and thus the total duration of the sequence, one reduces the spectral weight of the pulse sequence at the 1--2 transition. Moreover, gate error exhibits an oscillatory behavior, with minima corresponding to points where there is destructive interference at the leakage transition. For the $(\pi/2)_y$ pulse, fidelity of 99.9\% is achieved in 100 pulses, corresponding to a 20 ns gate time for a 5 GHz qubit, while for a $\pi$ pulse 99.9\% fidelity is achieved in around 300 pulses.

\section{Analysis of SFQ Gate Error}

In this Section, we present a detailed analytical treatment of gate errors due to the following three sources: (A) finite width of the SFQ pulses, (B) SFQ pulse timing jitter, and (C) leakage to higher energy levels of the qubit. We compare the ideal SFQ-based gate, where coherent rotations are realized from $\delta$-function pulses with no timing error and where the qubit is treated as an ideal two-level system, to the actual SFQ-based gate, where the pulses have finite width and timing jitter and where weak anharmonicity of the qubit is explicitly taken into account. We compute the state-averaged overlap fidelity of a qubit gate as follows \cite{Pedersen07}:
\begin{align}
F_{\rm avg}({\cal U}_{\rm id},{\cal U}_{G}) = \frac{2 + \left|{\rm Tr}({\cal U}_{\rm id}^\dag {\cal U}_{G})\right|^2}{6},
\label{eq:Fav}
\end{align}
where ${\cal U}_{\rm id}$ is unitary time evolution operator for the ideal gate and ${\cal U}_{G}$ corresponds to the actual gate. We will evaluate the fidelity of SFQ-based rotations by angle $\Theta$ about the $y$-axis, so we take
\begin{equation}
{\cal U}_{\rm id}=\exp\left(\frac{i\Theta \hat\sigma_y}{2}\right).
\label{eq:Uid}
\end{equation}
We compose this rotation from $n$ smaller rotations by angle $\delta \theta=\Theta/n$ about the $y$-axis, interspersed with appropriate free precession intervals that are matched to the Larmor period $2\pi/\omega_{10}$ of the qubit. The unitary operator describing the $\delta$-function pulses is given as follows:
\begin{equation}
\label{eq:Udelta}
{\cal U}_\delta^{(1)} = \exp\left(\frac{i\delta\theta\hat\sigma_y}{2}\right).
\end{equation}

Similarly, free precession for interval $t$ is described by the unitary operator
\begin{equation}
U_f(t)=\exp \left( \frac{i\omega_{10} t\hat\sigma_z}{2} \right).
\end{equation}

The actual evolution operator ${\cal U}_G$ is composed as a product of single-pulse evolution operators
${\cal U}_G^{(1)} $ and free  evolutions between pulses.  We assume that the SFQ pulse vanishes outside the time interval $(-t_c,t_c)$ and that the evolution during the pulse is defined by the differential equation
\begin{equation}
i\hbar \frac{\partial U_G^{(1)}(t)}{\partial t} = H(t) U_G^{(1)}(t),
\end{equation}
with the initial condition $U_G^{(1)}(-t_c)=1$.  The evolution operator at time $t_c$ thus defines the overall effect of a single pulse on the qubit state: ${\cal U}_G^{(1)}(\delta\theta)=U_G^{(1)}(t_c)$.

In the following, we consider the structure of the actual evolution operators ${\cal U}_{G}(\Theta)$ and calculate gate infidelity for three sources of error.

\subsection{Finite pulse width}

Here we analyze the effect of the finite SFQ pulse width. We begin by considering rectangular SFQ pulses with width $2t_c$.  The full Hamiltonian during the pulse is
\begin{align}
H=\frac{\hbar\omega_{10}}{2}(\eins-\hat\sigma_z)-\frac{\hbar\delta\theta}{4 t_c}\hat\sigma_y,
\end{align}
where $\delta \theta$ is the rotation angle induced by a single pulse. The corresponding evolution operator during the pulse represents precession in the field $(0, \delta\theta/2t_c,\omega_{10})$ and has the form
\begin{align}
{\cal U}_{\rm rect}^{(1)}=
\exp\big(i (2\omega_{10}t_c\hat\sigma_z+\delta\theta\hat\sigma_y)/2\big).
\end{align}

\begin{figure}
  \includegraphics[width=.45\textwidth]{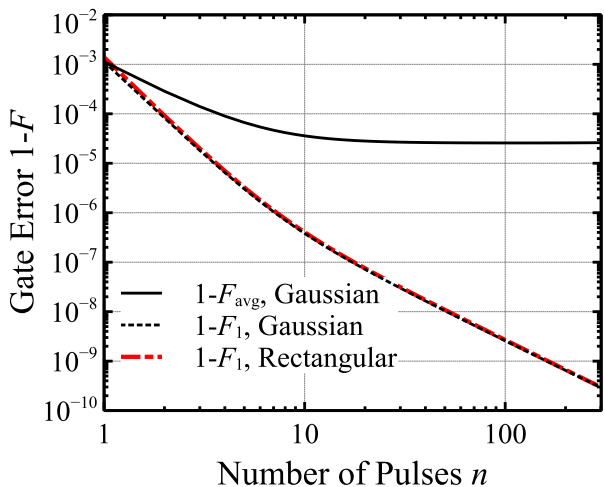}\\
  \caption{(Color online) Dependence of gate error on number of pulses $n$ used to realize a $(\pi/2)_y$ rotation for SFQ pulses of finite width.  The error $1-F_{1,rect}$ for a single rectangular pulse with width $2t_c$~=~7~ps was calculated from Eq.~\eqref{eq:rect} and is shown as the dash-dotted trace. The error for a single Gaussian pulse with width $\tau$~=~4~ps was computed numerically and is shown as the dotted trace. The error for the full $\pi/2$ rotation realized from $n$ Gaussian pulses is shown as the solid trace.
\label{figS1:finitewidth}}
\end{figure}

In the $\delta$-function approximation, the evolution during the same time interval would be
\begin{align}
{\cal U}_{\rm id}^{(1)}=\exp(i\omega_{10} t_c \hat\sigma_z/2){\cal U}_\delta^{(1)}\exp(i\omega_{10} t_c \hat\sigma_z/2).
\end{align}
Using Eq.~\eqref{eq:Fav}, we obtain the overlap error $1-F_{1,\rm rect}$ for a single pulse up to fourth order in $t_c$ and $\delta\theta$:
\begin{align}
1-F_{1,\rm rect}=\frac{1}{216}\left(
\delta\theta^4 \omega_{10}^2t_c^2+\delta\theta^2 \omega_{10}^4t_c^4
-\frac{\delta\theta^4 \omega_{10}^4t_c^4}{5}
\right).
\label{eq:rect}
\end{align}
This expression gives the important message that for short pulses, to the lowest order in $\omega_{10}t_c$,
the error decreases as $\delta\theta^4$ for decreasing $\delta\theta$.  However, for very small $\delta\theta \lesssim \omega_{10}t_c$, the error becomes quadratic in the rotation angle $\delta\theta$. We present the gate error due to rectangular pulses in Fig.~\ref{figS1:finitewidth} as the dash-dotted trace.

Next, we model the SFQ pulse by a Gaussian shape with width $\tau$:
\begin{align}
V(t) = \frac{\Phi_0}{\sqrt{2\pi}\tau} e^{-t^2/2\tau^2}.
\end{align}
The time-dependent Hamiltonian is given by
\begin{align}
H(t) &= \frac{\hbar\omega_{10}}{2}(\eins-\hat\sigma_z) + \frac{\hbar \, \delta \theta}{2\sqrt{2\pi} \tau} \,\, e^{-(t-t_k)^2/2\tau^2} \hat{\sigma}_y,
\end{align}
where $t_k$ is the arrival time of the SFQ pulse. We denote the time evolution operator for the full Gaussian pulse as ${\cal U}_{\rm Gauss}^{(1)}$, and we compute this operator over the interval $(-t_c,t_c)$, where we take $t_c = 5 \tau$. We assume that the SFQ pulse vanishes outside of the time interval $(-t_c,t_c)$ and that qubit evolution is described by the free evolution operator $U_f(2\pi/\omega_{10}-2t_c)$ during the time $2\pi/\omega_{10}-2t_c$.
The gate error for a single Gaussian SFQ pulse can be evaluated according to Eq.~\eqref{eq:Fav} and the result is shown in Fig.~\ref{figS1:finitewidth} as the dotted trace. The error for the Gaussian pulse closely follows the result for rectangular pulses with proper choice of $\tau$.

In addition, we have analyzed the fidelity of a gate composed of a resonant train of $n$ Gaussian SFQ pulses that is designed to realize a rotation by angle $\Theta = n \delta \theta$ about the $y$ axis.  The gate evolution operator is written as
\begin{align}
{\cal U}_G(\Theta)
 &= \left[U_{f}(2\pi/\omega_{10} - t_c){\cal U}_{\rm Gauss}^{(1)}(\delta\theta)
 U_{f}( - t_c) \right]^n\, .
\end{align}
Substituting this expression to Eq.~\eqref{eq:Fav}, we obtain the gate fidelity.  In Fig.~\ref{figS1:finitewidth}, we present the gate error as a function of the number of pulses $n$ for $\Theta=\pi/2$ and for Gaussian pulses with width $\tau$ = 4 ps. We observe that average gate error for a full rotation is $n^2$ times larger than the error of a single pulse, $1-F_{\rm avg} = n^2 (1-F_1)$.  For larger values of $n$, the single pulse error scales as $1/n^2$ [cf. Eq.~\eqref{eq:rect}] and the average gate error $1-F_{\rm avg} \propto (\omega_{10}\tau)^4 \Theta^2$ becomes independent of $n$, while remaining below  $10^{-4}$ due to the factor $(\omega_{10}\tau)^4$. Note that in these simulations we assume very long pulse times compared to what is achieved in practical SFQ circuits (where pulse widths $\tau<$~1~ps are readily accessible) in order to circumvent numerical errors associated with finite machine precision; the scaling of gate error with pulse duration can be understood from Eq.~\eqref{eq:rect}. For practical SFQ pulses, error associated with finite pulse duration is much smaller than the other two errors analyzed below.

\subsection{Pulse timing jitter}

\begin{figure}
  \includegraphics[width=0.45\textwidth]{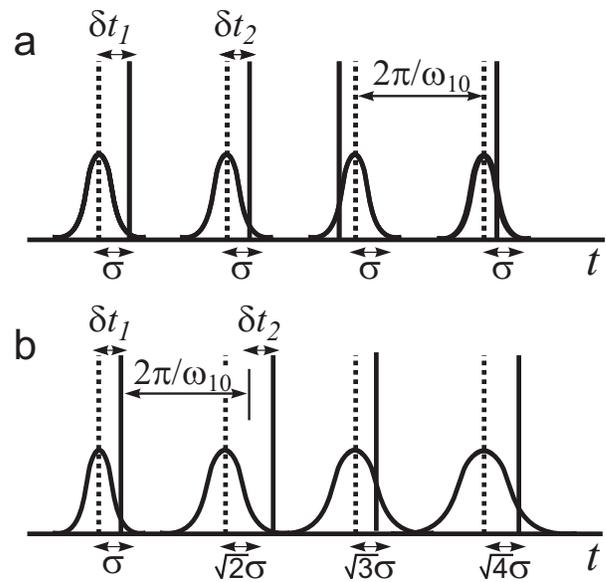}\\
  \caption{SFQ pulse timing jitter for two different clocking modes. Ideal pulses (dashed lines) are separated by an interval $2 \pi/\omega_{10}$. Arrival times of the actual pulses (solid lines) fluctuate due to timing jitter of the SFQ pulse generator. (a) For SFQ pulses triggered from a stable external clock, the timing jitter per pulse is constant. Pulse arrival times are normally distributed about the ideal pulse times with a constant width $\sigma$. (b) For SFQ pulses generated internally from a clock ring, timing errors accumulate incoherently, leading to a $\sqrt{k}$ degradation of timing jitter for the $k$th pulse.
\label{fig:SFQ_clocking}}
\end{figure}

Small variation in the arrival times of the SFQ pulses presents another source of gate error. As mentioned in Section II, the effect of timing jitter on SFQ gate fidelity depends on the manner in which the SFQ timing generator is triggered. We consider the following two cases: (1) \textit{External clock}. Here, the SFQ pulses are derived from a stable external clock, so that the timing error per pulse does not grow with the length of the sequence. (2) \textit{Internal clock}. Here, there is fixed error in the pulse-to-pulse spacing, so that errors in the timing of individual pulses accumulate incoherently as the length of the sequence grows. The effect of these two different clocking modes on pulse timing jitter is depicted schematically in Fig. \ref{fig:SFQ_clocking}. We examine these two cases in detail below.
For further discussion, we utilize an alternative expression to evaluate gate fidelity \cite{Bowdrey2002}:
\begin{align}
F_{\rm avg} = \frac{1}{6} \sum_{\alpha}F_{\alpha},\quad
F_\alpha={\rm Tr}\left\{
{\cal U}_G^{} \rho_\alpha {\cal U}_G^\dagger {\cal U}_{\rm id}^{} \rho_\alpha {\cal U}_{\rm id}^\dagger
\right\},
\label{eq:FavPure}
\end{align}
where the average is performed over the Pauli eigenstates $\rho_\alpha=|\alpha\rangle\langle \alpha|$ aligned along directions $\alpha=\pm x,\pm y,\pm z$.

\subsubsection{External clock}
We first analyze the effect of timing jitter on pulse trains derived from a stable external clock.  We assume that the pulse arrival times are distributed normally with respect to the external clock with distribution width $\sigma$.  For small jitter, $\omega_{10} \sigma \ll 1$, we can evaluate $F_{\alpha}$ using the following analysis.  The evolution of the qubit is characterized by a sequence of discrete rotations, Eq.~\eqref{eq:Udelta}, interspersed with intervals of free precession that are nominally matched to the qubit period $2 \pi/\omega_{10}$. Due to pulse timing jitter, the actual free precession interval between the $(k-1)$th and $k$th pulses becomes $2 \pi/\omega_{10}+\delta t_k-\delta t_{k-1}$, where $\delta t_k$ is the timing error associated with the $k$th pulse. For a qubit state vector that is initially aligned along the $z$-axis, the timing error causes the state to acquire a component $\delta y_k$ in the $y$-direction:
\begin{align}
\delta y_k=\omega_{10}(\delta t_k-\delta t_{k-1})\sin(k \delta\theta).
\label{eq:z_init}
\end{align}
Here, $k \delta \theta$ is the instantaneous polar angle of the qubit state vector. During the gate operation, the qubit state vector accumulates the error  $\delta Y=\sum_k \delta y_k$, and we find
\begin{align}
\label{eq:Fz1}
F_z & =1-\overline{\delta Y^2}/4, \\
\overline{\delta Y^2} & =(\omega_{10} \sigma)^2  \left[\sin^2(n\delta\theta)+ \delta\theta^2 \sum_{k=1}^{n-1} \cos^2(k\delta \theta )\right],
\nonumber
\end{align}
where the overbar represents an average over pulse jitter times $\delta t_k$. Assuming that $\delta\theta=\Theta/n$ is small, we can replace the summation by integration in the last expression and we find
\begin{equation}
F_{z} = 1- \left(\omega_{10}\sigma\right)^2 \left[\frac{\Theta^2}{8n} \left(1 + \frac{\sin{2 \Theta}}{2\Theta}\right) + \frac{\sin^2{\Theta}}{4}\right]
\end{equation}
for a qubit state initially aligned along the $z$-direction.

For a qubit state initially aligned along the $x$-axis, the analysis is the same with the replacement of $\sin(k\delta\theta)$ by $\cos(k\delta\theta)$ in Eq. \eqref{eq:z_init}. In this case we find
\begin{equation}
F_{x} = 1- \left[\frac{\Theta^2}{8n} \left(1 - \frac{\sin{2 \Theta}}{2\Theta}\right) + \frac{\cos^2{\Theta}}{4}\right]\, \left(\omega_{10}\sigma\right)^2.
\end{equation}
In the above expressions for $F_x$ and $F_z$ we have disregarded a small error along the $z$-direction, which is higher order in $\omega_{10}\sigma$.

In case of a qubit state vector initially aligned along the $y$-axis, the state vector remains close to the $y$-axis, and after each free precession acquires an error in the $x$-direction $\delta x_k\simeq \omega_{10}(\delta t_k-\delta t_{k-1})$.  This error is then rotated by the remaining $n-k$ pulses in $xz$-plane, resulting in the accumulation of total gate error along the $x$- and $z$-directions $\delta X=\sum_k \delta x_k \cos(\Theta-k\delta \theta)$ and
$\delta Z=\sum_k\delta x_k \sin(\Theta-k\delta \theta)$. For a qubit state initially aligned along the $y$-axis, we find a gate fidelity
\begin{align}
F_y = 1-\frac{\overline{\delta X^2}}{4}-\frac{\overline{\delta Z^2}}{4}.
\end{align}
Evaluating the summations for $\delta X$ and $\delta Z$ under the assumption of uncorrelated $\delta t_k$, we obtain
\begin{equation}
F_y=1-\frac{(\omega_{10} \sigma)^2}{4}\left(\frac{\Theta^2}{n}+1\right).
\end{equation}
The average gate error is computed from Eq.~\eqref{eq:FavPure}, and we find
\begin{equation}
\label{eq:Favgjitter}
F_{\rm avg}=1-\frac{(\omega_{10} \sigma)^2}{6}\left(\frac{\Theta^2}{n}+1\right).
\end{equation}

\subsubsection{Internal clock}

Next, we evaluate gate fidelity for a system where the SFQ pulses are clocked internally in such a way that the time interval between pulses fluctuates independently, so that error in the arrival times of individual pulses accumulates incoherently. The free evolution is determined by the time interval $2 \pi/\omega_{10}+\delta t_k$, where $\delta t_k$ is normally distributed and uncorrelated from pulse to pulse. Due to the timing error, a qubit state vector initially aligned along the $z$-direction acquires a spurious component $\delta y_k = \delta t_k\sin(k\delta\theta)$ along the $y$-axis.  We thus find $\overline{\delta Y^2}  =(\omega_{10} \sigma)^2  \sum_k \sin^2(k\delta \theta )$. Following the same procedure described in the previous section, we obtain a gate fidelity 
\begin{align}
F_z & =1-\frac{n (\omega_{10} \sigma)^2}{8}  \left[1-\frac{\sin 2\Theta}{2\Theta}
\right].
\end{align}
For a pure state initially aligned along the $x$-axis, we find
\begin{align}
F_x & =1-\frac{n (\omega_{10} \sigma)^2}{8}  \left[1+\frac{\sin 2\Theta}{2\Theta}
\right].
\end{align}

For states initially aligned along the $y$-axis, error accumulates along both the $x$- and $z$-directions, as discussed in the previous section. Evaluating the corresponding gate errors $\overline{\delta X^2}$ and $\overline{\delta Z^2}$,  we find
\begin{align}
F_y = 1-\frac{\overline{X^2}}{4}-\frac{\overline{Z^2}}{4}\simeq 1-\frac{n (\omega_{10} \sigma)^2}{4}.
\end{align}
The gate fidelity averaged over all qubit states is given by
\begin{align}
F_{\rm avg} = 1-\frac{n (\omega_{10} \sigma)^2}{6}.
\end{align}

\begin{figure}
  \includegraphics[width=0.45\textwidth]{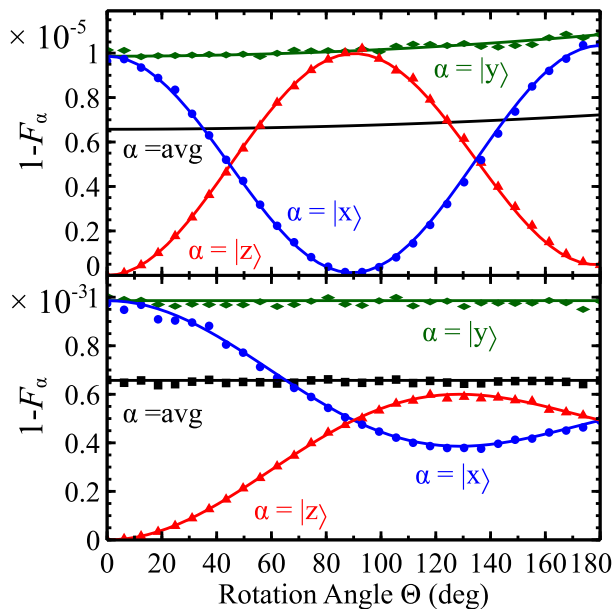}\\
  \caption{(Color online) Dependence of gate error due to pulse timing jitter on rotation angle $\Theta$ for SFQ pulses generated by an external clock (top panel) and an internal clock (bottom panel).  Here the rotation is realized from $n=100$ pulses and the standard deviation of pulse timing jitter is $\sigma=0.2$~ps.  Solid lines represent the analytical expressions for $F_\alpha$, while the points were obtained from numerical simulations of $10^4$ realizations of pulse timing jitter.
\label{figS2:jitter}}
\end{figure}

We have numerically evaluated SFQ gate error in the presence of timing jitter as a function of rotation angle $\Theta$ for pure initial states aligned along directions $\alpha=\{x,y,z\}$. Here we take $\sigma=0.2$ ps and $n=100$. For a given realization of timing jitter $\{\delta t_k\}$, we calculate the overlap of the final qubit state with the corresponding state obtained by the ideal gate, Eq.~\eqref{eq:Uid}, and then we average the overlap over $10^4$ realizations of $\{\delta t_k\}$.  The results are shown in the upper and lower panels of Fig.~\ref{figS2:jitter} for external and internal gating of the SFQ pulses, respectively. The simulation results are plotted as points, and the lines represent the analytical expressions derived above.

\subsection{Leakage to higher energy levels of the qubit}

Finally, we analyze the effect of weak qubit anharmonicity on SFQ gate fidelity.  We treat the qubit as a three-level system with anharmonicity $\eta = (\omega_{10}-\omega_{21})/\omega_{10}$. The Hamiltonian is given by Eqs. (13) and (14).  The corresponding time evolution operator is a three-dimensional unitary matrix and the definition for the average  fidelity has to be modified accordingly.  However, since we are interested in averaging over the two-level qubit subspace of the system Hilbert space, the average fidelity reduces to \cite{Pedersen07}
\begin{align}
F_{\rm avg}({\cal U}_{\rm id},{\cal U}_{G}) = \frac{{\rm Tr}\{{\cal U}_{G}^\dag {\cal P} {\cal U}_{G}^{} {\cal P}\} + \left|{\rm Tr}\{ {\cal P} {\cal U}_{\rm id}^\dag {\cal U}_{G}\}\right|^2}{6},
\label{eq:Fav3}
\end{align}
where ${\cal P}$ is the projection operator on the qubit subspace. This expression for fidelity is consistent with Eq.~\eqref{eq:FavPure} provided we use the following modified three-dimensional unitary operator to describe evolution under the ideal gate:
\begin{equation}
{\cal U}_{\rm id} =
\left(
   \begin{array}{ccc} 
      \displaystyle \cos(\Theta/2) &
      \displaystyle  \sin(\Theta/2) & 0 \\
      \displaystyle  -\sin(\Theta/2) &
      \displaystyle  \cos(\Theta/2) & 0\\
      0 & 0 & 1
   \end{array}
\right).
\end{equation}

\begin{figure}
  \includegraphics[width=0.45\textwidth]{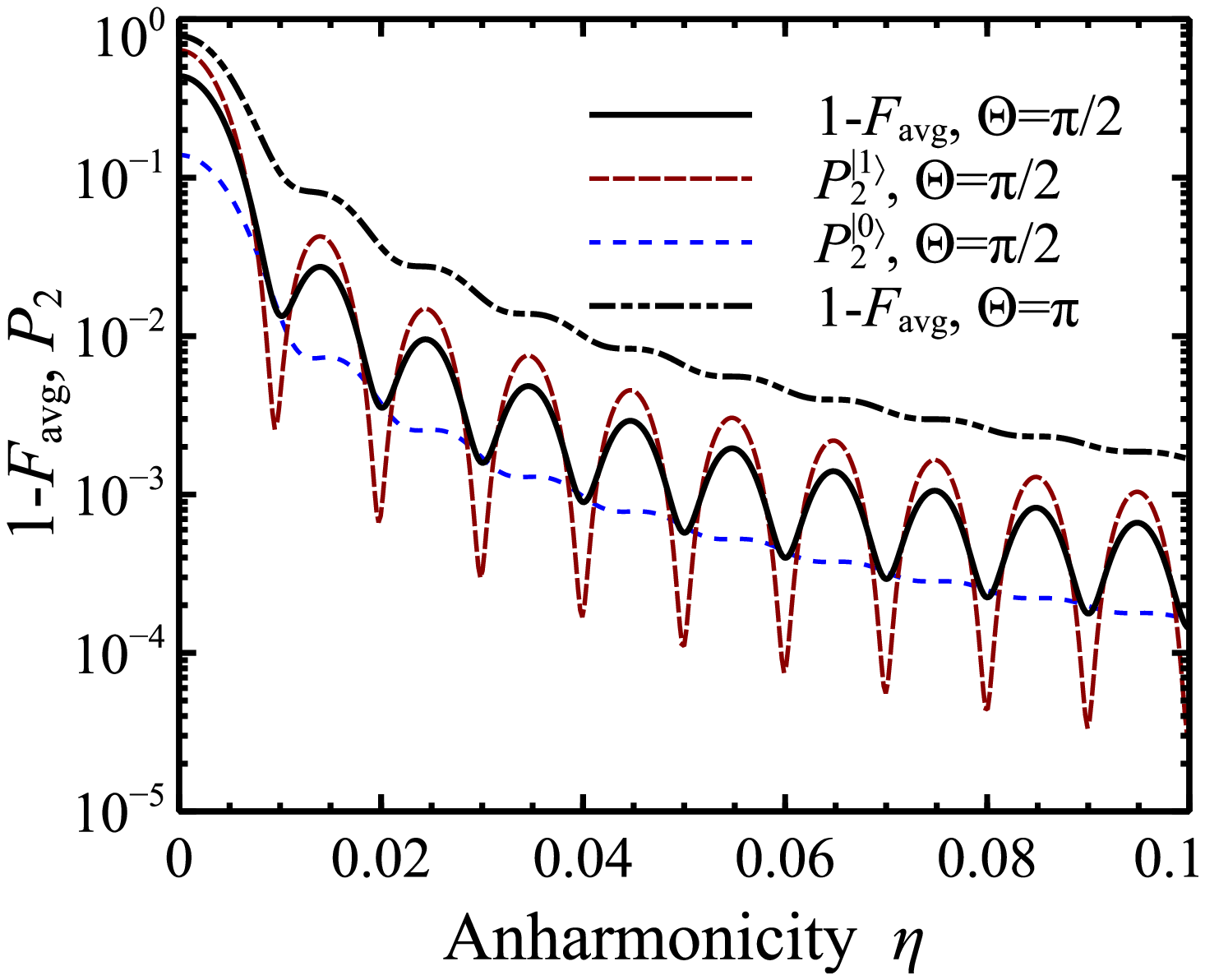}\\
  \caption{(Color online) Dependence of gate error on qubit anharmonicity $\eta$ for $(\pi/2)_y$ (solid line) and $\pi_y$ (dash-double-dot line) rotations.  For comparison, the occupation probability of the $\ket{2}$ state is shown for initial ground, $P_2^{|0\rangle}$, and excited, $P_2^{|1\rangle}$, qubit states in the case of the $(\pi/2)_y$ rotation.  The average error always exceeds the smallest of $P_2^{|0\rangle}$ and  $P_2^{|1\rangle}$.  The number of pulses to perform the rotation is $n=100$. The presented curves $P_2^{|0\rangle}$ and  $P_2^{|1\rangle}$ were obtained by numerically solving the evolution of the initial ground and excited states under the SFQ pulse train, but these curves are nearly indistinguishable from those obtained from analytical expression \eqref{eq:P2pm} for $\eta\gtrsim 0.01$ (not shown).
\label{figS3:jitter}}
\end{figure}

We evaluate the error of a $\Theta_y$ gate due to the presence of the second excited state by summing the spurious amplitude of the $\ket{2}$ state induced by pulse $k$ as
$\delta\psi^{|j\rangle}_{2,k}=\exp(2 \pi i\eta(n-k))
(\delta\theta/\sqrt{2})\psi^{|j\rangle}_{1,k-1}$,
where $\psi^{|j\rangle}_{1,k-1}$ is the probability amplitude of the qubit being in the first excited state at time of pulse $k$ if it was initially in state $\ket{j}$ with $j=0,1$. Here the factor $\exp(2\pi i\eta(n-k))$ represents the phase acquired by the second excited state over the remainder of the sequence following the $k$th pulse.
Performing summation over $n$ SFQ pulses, we obtain the probability of excitation to the second excited state as
\begin{equation}
P_2^{|j\rangle} = \frac{\Theta^2}{8n^2}\left|
\frac{1-e^{in(2\pi\eta+\delta\theta/2)}}{1-e^{i(2\pi\eta+\delta\theta/2)}}-(-1)^j
\frac{1-e^{in(2\pi\eta-\delta\theta/2)}}{1-e^{i(2\pi\eta-\delta\theta/2)}}
\right|^2.
\label{eq:P2pm}
\end{equation}
Here we have assumed that the $\ket{1}$ state amplitudes \\$\psi_{1,k}^{|0\rangle}=\sin (k \delta\theta/2)$ and  $\psi_{1,k}^{|1\rangle}=\cos (k \delta\theta/2)$ are not significantly modified by the small amount of leakage to the second excited state, and we have disregarded direct ${|0\rangle}\to {|2\rangle}$ transitions.
The numerically evaluated curves for $P_2$ in Fig. \ref{fig:fig3} are well described by Eq.~\eqref{eq:P2pm} for $n\gtrsim 10$. In particular, the fidelity decreases as $n^{-2}$ for large $n$, in addition to displaying an oscillating component that is more pronounced for smaller gate rotation angle $\Theta$. In Fig.~\ref{figS3:jitter} we present average gate error as a function of anharmonicity $\eta$ for $(\pi/2)_y$ and $\pi_y$ gates realized using $n=100$ SFQ pulses. The infidelity drastically decreases for $\eta\gtrsim 1/n$ and then exhibits a slower decrease with a minimum at $\eta=1/2$.  The oscillations of $1-F_{\rm avg}$ have the period $\Delta \eta\sim 1/n$ and nearly disappear for a $\pi_y$ rotation.  In this figure, we also plot the $\ket{2}$ state occupation $P^{\ket{j}}_2$ following the $(\pi/2)_y$ rotation for the qubit initially in state $\ket{j}$.  Comparison of our numerical calculations of $P_{2}^{\ket{j}}$ with Eq.~\eqref{eq:P2pm} shows that the two agree well for $|\eta|\gtrsim 1/n$.

\section{Conclusion}

Our simulations thus indicate that errors due to SFQ timing jitter and weak qubit anharmonicity are roughly of the same order $\sim 10^{-3}$ for SFQ pulse trains consisting of around 100 pulses, corresponding to 20 ns for a $\pi/2$ rotation of a 5 GHz qubit. While these errors are nonnegligible, they are nevertheless small enough to enable robust qubit control with fast gates at error levels below the threshold for a fault-tolerant superconducting surface code \cite{Fowler12}. Gate errors could be suppressed further by efforts to improve the timing stability of the SFQ circuit or by simple circuit redesign to increase qubit anharmonicity. Here we have attempted to analyze only the simplest SFQ pulse trains. State-of-the-art SFQ timing generators should allow the realization of robust sequences with arbitrary interpulse delays. We anticipate that optimal control tools of the sort used to optimize microwave-based single-qubit gates \cite{Motzoi09} and fast two-qubit gates \cite{Motzoi11} could also be employed to engineer SFQ sequences with interpulse delays designed to suppress $\ket{2}$ state errors and increase gate speed and/or fidelity over the naive gate sequences considered here.

Due to technical complexities of transmitting SFQ pulses from chip to chip, the practical realization of SFQ-based qubit gates will require the on-chip integration of the qubit circuit with at least a handful of SFQ elements. While in the past the high static dissipation of SFQ circuits has presented an obstacle to millikelvin-temperature operation, the recent development of low-power biasing schemes for reciprocal quantum logic (RQL) \cite{Herr11} and energy efficient SFQ logic (eSFQ) \cite{Mukhanov11} opens the door to the integration of SFQ and qubit circuits on the same chip. Care must be taken to isolate the qubit circuit from nonequilibrium quasiparticles generated in the SFQ control circuit; however, quasiparticle poisoning of the qubit circuit can be mitigated by avoiding direct galvanic connection between the signal and ground traces of the SFQ and qubit circuits. The ability to generate fluxons in close proximity to the qubit circuit will provide a high degree of robustness to the SFQ-based rotations, due to the quantization of flux associated with the SFQ pulses.

In conclusion, we have described a method for the high-fidelity coherent manipulation of superconducting qubit and linear cavity modes using resonant trains of SFQ pulses. The SFQ pulse trains can be generated locally in the qubit cryostat without the need for an external microwave generator. Taken together with a recent proposal to map the quantum information in a cQED circuit to a binary digital output using a Josephson microwave photon counter \cite{Govia14}, this work points a direction toward the integration of large scale superconducting quantum circuits with cold control and measurement circuitry based on SFQ digital logic.

\vspace{12pt}

\begin{acknowledgments}
We thank B. L. T. Plourde for helpful discussions.
\end{acknowledgments}

\end{document}